\documentclass[12pt,aps,floats,showpacs,amssymb,tightenlines]{revtex4}

\usepackage{amsmath}
\usepackage{amsfonts}
\usepackage{amscd}
\usepackage{epsfig}
\usepackage{amssymb}
\usepackage{tabularx}

\newcommand{\e}{\epsilon}
\newcommand{\m}{\mu}

\renewcommand{\r}{\rho}
\newcommand{\w}{\omega}

\newcommand{\p}{\partial}

\newcommand{\s}{ Schwarzschild }
\renewcommand{\o}{{\cal{O}}}

\begin{document}

\title{Instability of the negative mass Schwarzschild naked singularity}
\author{Reinaldo J. Gleiser and Gustavo Dotti}
\affiliation{Facultad de Matem\'atica, Astronom\'{\i}a y F\'{\i}sica,
Universidad Nacional de C\'ordoba, Ciudad Universitaria,
(5000) C\'ordoba, Argentina}
\email{gdotti@famaf.unc.edu.ar}

\begin{abstract}
We study the negative mass  Schwarzschild spacetime, which has a
naked singularity, and show that it is perturbatively unstable. This
is achieved by first introducing a modification of the well known Regge - Wheeler - Zerilli approach to black hole perturbations to allow for the presence of a ``kinematic'' singularity that arises for negative masses, and then exhibiting exact exponentially growing  solutions to the
linearized Einstein's equations. The perturbations are smooth everywhere and behave nicely around the
singularity and at infinity. In particular, the first order variation of the
scalar invariants can be made everywhere arbitrarily small as compared to the zeroth order terms.
Our approach is also compared to a recent analysis that leads to a different conclusion regarding the stability of the negative mass  Schwarzschild spacetime. We also comment on the relevance of our results to the stability of more general negative mass, nakedly singular spacetimes.
\end{abstract}

\maketitle

\noindent
\section{Introduction: the negative mass Schwarzschild solution}

The positive mass theorems of general relativity establish that, under plausible assumptions,
asymptotically flat solutions of the Einstein equations with physically acceptable matter sources
cannot have negative total mass. Another well known feature of general relativity, compatible with
the positivity theorems, is the possibility of the presence, or eventual formation, of singularities
 in the spacetime manifold. However, in the most relevant examples where this happens, the singularity
 is surrounded by a horizon, so that it is unobservable to an observer outside the horizon. That this
 should happen generically in all physically relevant situations is the essential content of the versions
 of the Cosmic Censorship hypothesis. On the other hand, from a strictly mathematical point of view,
 there is plenty of solutions of the Einstein equations that are asymptotically flat and present a
  negative total mass, but, in general, they are also nakedly singular, which would make them physically
   unacceptable. A simple and particularly interesting example of this type of solutions is the
   negative mass Schwarzschild spacetime. Nevertheless, for different reasons, these solutions
   are not expected to arise from physically acceptable regular initial data. One should notice,
   however, that what is physically acceptable may change as more data is available. What we
   have in mind here is the recent suggestion of the existence of ``phantom matter", that is some
   field that carries {\em negative kinetic energy}, as a possible explanation for the observed
    acceleration of the scale factor of the Universe \cite{pm}.

A different, but equally important consideration regarding negative masses is the manner
in which they interact with positive masses. This gives rise to different types of
instabilities that have been analyzed in the literature. We refer to \cite{ghi} for a
recent review. It is rather remarkable, however, that until recently \cite{ghi}, the stability
of the negative mass Schwarzschild solution, that, in a loose sense, may be considered as representing
the field of an isolated negative mass source, had not been considered in the literature, although the
 essential machinery for this analysis, at least at the linear level, had been available for some time.
 The analysis carried out in \cite{ghi} leads to the conclusion that if certain boundary conditions,
 assumed suitable for the problem, are imposed, the negative mass Schwarzschild singularity should be considered as
 stable under linear perturbations. However, as discussed in Section \ref{sean}, the marginally stable  solution of
  Zerilli's equation used to prove stability in
 \cite{ghi}, is only a twice differentiable functions  of the radial coordinate. As a consequence, the
associated  perturbed metric is not continuous and thus it is not a
solution of the linearized vacuum  Einstein's
 equations (it
 actually satisfies the field equations of a singular distribution of
 sources).
  Given this situation, we undertake in this paper
 a new analysis of the linear stability of the negative mass Schwarzschild solution. This is based on
 a modification of the standard Regge-Wheeler-Zerilli formalism (reviewed in Section \ref{rw})
  that simplifies and makes more
 transparent the analysis of the perturbative equations for negative mass. This is given in Section
 \ref{rwzg}.
  Next, we consider the existence of ``zero modes'', i.e., static solutions of the perturbation equations.
   We give the general solution, that for each mode depends on two integration constants, and prove that
    there is no choice of these constants such that the solution can be considered as a ``small"
    perturbation. We give also a physical interpretation for these solutions that supports our
    conclusions. In Section \ref{us1}, based on this modified form of the formalism, and this is the
     main result presented in this paper, we prove that under quite general boundary conditions,
      {\em all} scalar perturbative modes are unstable. Our results and possible consequences are
      finally summarized in Section \ref{cc}.

\section{Perturbation equations for the spherically symmetric Schwarzschild solution.} \label{rw}

The vacuum solution of Einstein's equations for the exterior of a bounded spherically symmetric source,
first obtained by Schwarzschild, may be written in the form,
\begin{equation}
\label{Schw1}
ds^2 =  -\left(1-\frac{2 M}{r}\right) dt^2 +
 \left(1-\frac{2 M}{r}\right)^{-1} dr^2   +r^2 \left( d \theta^2 +  \sin^2(\theta)
d \phi^2 \right)
\end{equation}
where $\theta$ and $\phi$ are standard coordinates on the unit sphere, and $M$ is a real
(positive or negative) constant, that can be identified with the Newtonian mass of the source.
The range of $r$ is limited by the singularities of the metric coefficients in (\ref{Schw1}).
 For $M>0$ the singularity at $r=2M$ corresponds to a regular horizon, and the solution may be
extended to $r<2M$. The resulting space time is known as a ``black
hole''. The horizon hides
  a curvature singularity for $r=0$, rendering the portion of space time outside the horizon
  globally hyperbolic. Similarly, for negative mass ($M<0$) there is a singularity for $r=0$,
   but in this case there is no horizon, the singularity is ``naked'', and the resulting space
    time is not globally hyperbolic.

The rather strange properties of the spacetime described by
(\ref{Schw1}) led to question the possibility of their physical
realization. Since the solution (at least for $r>2M$) is static, a
first step in this direction, carried out by Regge and Wheeler \cite{RW}, and
completed later by Zerilli \cite{Zerilli}, was aimed at establishing their
stability, through an analysis of the behavior of linear
perturbations. The Einstein equations lead in this case to a set of
coupled linear partial differential equations for the functions
representing the perturbations. Because of the spherical symmetry of
(\ref{Schw1}) this functions may be grouped into ``even'' (or
scalar) and ``odd'' (or vector) sets that do not couple to each
other, and are characterized by their dependence on the angular
coordinates $(\theta,\phi)$. Still, the resulting perturbative
equations contain, in general, some redundant information that can
be eliminated through a choice of ``gauge''.
A particular choice, that keeps full generality, was selected in
\cite{RW}, \cite{Zerilli} and is known as the ''Regge-Wheeler gauge''. Independently of the gauge issue, the linear stability of the Schwarzschild space time, well known for $M>0$, was also established for vector perturbations for $M<0$ in \cite{ghi}. For this reason we shall restrict our analysis to the scalar
perturbations.  In the Regge-Wheeler gauge, the scalar (even)
perturbations, for each angular mode $(\ell,m)$,  are fully
described by  four functions $H_0(r,t)$, $H_1(r,t)$, $H_2(r,t)$ and
$K(r,t)$, in terms of which the perturbed metric takes the form,
\begin{eqnarray}
\label{RW1}
ds^2 & = & -\left(1-\frac{2M}{r}\right)\left(1- \e H_0 Y_{\ell,m}\right)dt^2 +
2 \: \e \;  H_1 Y_{\ell,m} dt dr+\left(1-\frac{2M}{r}\right)^{-1}\left(1+ \e H_2 Y_{\ell,m}\right)dr^2
\nonumber \\
& & +r^2\left(1+ \e K\, Y_{\ell,m}\right) \left( d \theta^2 +  \sin^2(\theta)
d \phi^2 \right)
\end{eqnarray}
where $\epsilon$ is an auxiliary parameter, useful in keeping track of the order of the perturbation.
$Y_{\ell,m}=  Y_{\ell,m}(\theta,\phi)$ are standard spherical
harmonics on the sphere, and, on account of the spherical symmetry, we may restrict to $m=0$ without loss of generality. The linearized Einstein equations for the
metric (\ref{RW1}) imply $H_0(r,t)=H_2(r,t)$, and a set of coupled
differential equations for $H_1$, $H_2$ and $K$. This can be solved
in terms of the Zerilli function $\psi_Z(r,t)$, by the replacements,
\begin{eqnarray}
\label{RW2}
K & = & f(r) \psi_Z+\left(1-\frac{2M}{r}\right)\frac{\partial \psi_Z}{\partial r} \nonumber \\
H_1 & = & h(r)\frac{\partial \psi_Z}{\partial t}+r\frac{\partial^2 \psi_Z}{\partial t \partial r}
 \nonumber \\
H_2 & = & \frac{\partial }{\partial r}\left[\left(1-\frac{2M}{r}\right)\left(h(r) \psi_Z
+r\frac{\partial \psi_Z}{ \partial r} \right)\right] -K
\end{eqnarray}
where,
\begin{eqnarray}
\label{RW2a}
f(r) & = &  \frac{\lambda(\lambda+1)r^2+3 \lambda M r + 6 M^2}{r^2(\lambda r + 3 M)}\nonumber \\
h(r) & = &  \frac{\lambda r^2 - 3 \lambda r M -3 M^2}{(r-2M)(\lambda r + 3M)}\nonumber \\
\lambda & = &  \frac{(\ell-1)(\ell+2)}{2}
\end{eqnarray}
provided $\psi_Z$ is a solution of the Zerilli equation,
\begin{equation}
\label{RW3}
\frac{\partial^2 \psi_Z}{\partial t^2}-\frac{\partial^2 \psi_Z}{\partial r_*^2} +V(r) \psi_z = 0
\end{equation}
where,
\begin{equation}
\label{poten1}
V(r) = 2 \left(1 -\frac{2 M}{ r}\right) {\frac{ \lambda^2 r^2
\left[(\lambda+1) r + 3 M \right] + 9 M^2 (\lambda r +M)}{ r^3
(\lambda r + 3 M)^2} } ,
\end{equation}
and $r_*$ is the ``tortoise'' coordinate, related to $r$ by,
\begin{equation}
\label{RW4}
\frac{dr_*}{dr}=\left(1-\frac{2M}{r}\right)^{-1}
\end{equation}

In what follows, for simplicity, we choose the integration constant relating $r$ and $r_*$ such that,
\begin{equation}
\label{RW4a}
r_* = r +2 M \left[\ln (r-2M) -\ln(2|M|)\right]
\end{equation}
Therefore, for $M<0$, we have $ r_*=0$ for $r=0$.

We notice that the relations (\ref{RW2}) can be inverted, giving,
\begin{equation}
\label{RW3a}
\psi_Z(r,t)  = \frac {r (r-2M)}{ ( \lambda+1 )  ( \lambda
\,r+3\,M ) } \left(H_2
 - r  \frac {\partial K}{\partial r}
\right)
    +\frac{r}{\lambda+1} K
\end{equation}
Therefore, there is a one to one correspondence between $\psi_Z$ and
the Regge-Wheeler gauge functions, $H_0$, $H_1$, $H_2$ and $K$.

For the stability analysis one considers a time dependence of the form,
\begin{equation}
\label{psitopsi1}
\psi_Z(t,r)=e^{i \omega t} \psi(r)
\end{equation}
which turns (\ref{RW3}) into the Schr\"odinger-like form,
\begin{equation}
\label{RW3b}
 -\frac{d^2 \psi_Z}{d r_*^2} +V(r) \psi = \omega^2  \psi
\end{equation}
Very loosely speaking, stability is attained if (\ref{RW3b}) has no solutions with {\em negative}
 $\omega^2$. More rigorously, this statement must be supplemented with the imposition of appropriate
 boundary conditions on $\psi(r)$. The resulting boundary value problem for $M > 0$ turns out to be
  quite different from that for $M<0$.

The case $M >0$ has been considered in detail in the literature
(see, e.g. \cite{chandra}). In this case we have $-\infty < r_* <
+\infty$ and $V(r)$ is positive definite in this range. Then,
 imposing boundary conditions that imply that the perturbed space time remains asymptotically flat,
  with a regular horizon, it can be shown that (\ref{RW3b})is turned into a self adjoint boundary
  value problem, that admits only solutions with $\omega^2 >0$, and, therefore, the spacetime is
  linearly stable.

For $M<0$ the problem is quite different. The domain in this case is
 $0 \leq r_* < +\infty$ and, besides the singularity of $V$ in (\ref{poten1})at
 $r=r_*=0$, there is another one at the interior point
 $r = -3 M/\lambda$. Also,
 it is not at all clear what boundary conditions are physically adequate
 for $r_*=0$. Finally, Zerilli's auxiliary function $\psi_Z(t,r)$ is generically
 singular (at $r = -3 M/\lambda$) for smooth metric perturbations.
  An attempt at an appropriate formulation of the problem was recently
 given in \cite{ghi}.
  Unfortunately, for reasons explained below in Section \ref{sean}, we consider
 that there are some difficulties in that analysis that require a revision of their conclusions.

\section{A modified Regge-Wheeler-Zerilli formalism}
\label{rwzg}

In view of the difficulties indicated in the previous Section, we
present here a slight modification of the Regge-Wheeler-Zerilli
formalism that, as we shall show, allows for a simple analysis of
 the stability problem for $M<0$, with the intriguing result
 that these solutions are unstable
 under arbitrarily small perturbations.  Following \cite{ghi}, we define
  $\mu = -2M$, but consider,
 instead of $\psi$, the function,
\begin{equation}
\label{AppB1} \zeta(t,r)= (2\lambda r -3 \mu ) \psi_Z(t,r).
\end{equation}
 { } From (\ref{RW3}) we easily find that $\zeta$ satisfies the equation,
 \begin{equation} \label{RW17}
 {\frac {\partial ^{2} \zeta}{\partial {t}^{2}}} = {\cal O}
 \zeta, \end{equation}
 where
 \begin{multline}
 {\cal O} \zeta \equiv
{\frac {   \left( r+\mu \right) ^{2}}{{r}^{2}}}\left( {\frac
{\partial^{2} \zeta}{\partial {r}^{2}}}  \right)   -{\frac {\left(
4\lambda {r}^{2}+ 6 \lambda \mu r-3{\mu}^{2} \right) \left( r+\mu
 \right)  }{{r}^
{3} \left( 2 \lambda r-3\mu \right) }}{\frac {\partial
\zeta}{\partial r}} \\  -{\frac {
 \left( r+\mu \right)  \left( 3{\mu}^{2}- 6 \lambda \mu r+4 \lambda^2 {r}^{2}
 \right) }{{r}^{4} \left( 2 \lambda \,r-3\,\mu \right) }}\zeta
  \nonumber
\end{multline}

Note that $\zeta(t,r)$ is also simply related to $H_1$, $H_2$, and
$K$:
\begin{eqnarray} \nonumber
 K &=& \frac{(\mu+r) r \left( \frac{\p \zeta}{\p r}\right) + (\lambda r - \mu) \zeta}{r^2 (2\lambda r - 3
 \mu)}\\ \label{nv1}
H_1 & = & \frac{2 r (\mu+r) \left( \frac{\p^2 \zeta}{\p t \p r}
\right) + \mu \left( \frac{\p \zeta}{\p t} \right)}{r (r+\mu) (2
\lambda r - 3 \mu)} \\ \nonumber H_2 & = & \frac{(r+\mu)}{(2 \lambda
r - 3 \mu)} \left( \frac{\p^2\zeta}{\p r^2} \right) -  \frac{4
\lambda r^2 + 6 \lambda \mu r - 3 \mu^2}{2 r (2 \lambda r - 3
\mu)^2} \left( \frac{\p \zeta}{\p r} \right) \\ && + \frac{3 \mu^2 -
6\lambda \mu r + 4 \lambda^2r^2}{2 r^2(2 \lambda r - 3 \mu)^2}
\zeta(t,r) \nonumber
\end{eqnarray}
and shares with $\psi_Z(t,r)$ its uniqueness and gauge
 invariance properties.
We may regard (\ref{nv1}) as a map $\Psi:= \zeta \to (K,H_1,H_2)$, a
possible inverse map being
\begin{equation} \label{im}
\zeta(t,r) = -\left(\frac{2 r^2 (r + \mu)}{\lambda+1} \right)
\frac{\p K}{\p r} + \left( \frac{(2 \lambda r-3 \mu) r}{\lambda+1}
\right) K + \left(\frac{2 r (r + \mu)}{\lambda +1} \right) H_2
\end{equation}
The following results follow straightforwardly from the definition
of $\Psi$, eqn (\ref{nv1}):\\

\noindent {\em Lemma 1:}  the metric perturbation given by
$\Psi(\zeta(t,r))$ is $C^{\infty}$ (for $r \in (0,\infty)$) if and
only if $\zeta(t,r)$ is $C^{\infty}$ and satisfies the condition
\begin{equation} \label{cz}
\left. \frac{\p \zeta(t,r)}{\p r}  \right|_{r=3\mu/(2 \lambda)} =
\left. \frac{2 \lambda^2}{3 \mu (2 \lambda+3)} \zeta(t,r)
\right|_{r=3\mu/(2 \lambda)}
\end{equation}
If (\ref{RW17}) is solved for initial data $\zeta(t=t_o,r)$ and $\p
\zeta /\p t (t=t_o,r)$ both satisfying (\ref{cz}), the solution will
satisfy  (\ref{cz}) for all $t$ where it is defined. In other words,
(\ref{cz}) is preserved by the evolution equation (\ref{RW17})
\\
{\em Proof:} if the metric perturbation $(K,H_1,H_2)$ is
$C^{\infty}$ then so is $\zeta$, as follows from (\ref{im}), and  we
can use Taylor's theorem in (\ref{nv1}) to prove (\ref{cz}) is
required to assure that $(K,H_1,H_2)$ are smooth at $r=3 \mu /(2
\lambda)$. The preservation of this constraint by the evolution
equation (\ref{RW17}) follows from the fact that the operator ${\cal
O}$ defined in (\ref{RW17}) sends functions satisfying (\ref{cz}) to
functions satisfying (\ref{cz}) $\Box$\\

Since the coefficients in (\ref{RW17}) depend only on $r$, we may
separate variables by setting $\zeta(t,r)=\exp(i \omega t) \chi(r)$,
which gives
\begin{eqnarray}
\label{AppB2} {\frac {d^2 \chi}{d{r}^{2}}} & = &
 {\frac { \left(
4 \lambda {r}^{2}+6  \lambda r\mu -3{\mu}^{2} \right)  }{
 \left( 2  \lambda  r-3\mu \right)  \left( r+\mu
 \right) r}}{\frac {d \chi}{dr}} \nonumber \\ & &
+\left[{\frac { \left( 4{\lambda}^{2}{r}^{2}-6 \lambda
r\mu+3{\mu}^{2} \right)
  }{ \left( 2  \lambda  r-3\mu \right)
 \left( r+\mu \right) {r}^{2}}}  -\frac{r^2}{(r+\mu)^2}\omega^2 \right]{\chi}
\end{eqnarray}
This ODE, obviously, does not have the Schr\"odinger like form of
(\ref{RW3b}). Moreover, the coefficients of $\chi$ and $d \chi/dr$
are singular at $r=3 \mu /  (2 \lambda)$. However, the nature of
this singularity is rather subtle, and, as shown below, {\em all}
solutions of (\ref{AppB2}) are regular around $r=3\mu /(2 \lambda)
$, {\em irrespective} of the
boundary conditions for $r=0$ and $r=\infty$.\\

\noindent {\em Lemma 2:} For any $\lambda,\mu$ and $\omega$, all
solutions of
(\ref{RW17}) are regular at $r=3 \mu /  (2 \lambda)$.\\
{\em Proof:} Let $\rho = r- 3 \mu / (2 \lambda)$, then expanding
(\ref{AppB2}) around $\r=0$ we get
   \begin{multline} \label{exp}
0 = \frac{d^2 \chi}{d \rho^2} + \left( -2\,{\rho}^{-1}-{\frac
{4{\lambda}^{2}}{3 \mu\, \left( 3+2\,\lambda
 \right) }}+\,{\frac {16 \left( 3+\lambda \right) {
\lambda}^{3}\rho}{9{\mu}^{2} \left( 3+2 \lambda
 \right)^2 }} + {\cal O} (\rho^2) \right) \frac{d \chi}{d \rho} \\+
 \left({\frac {-4{\lambda}^{2}}{3 \mu\, \left( 3+2\,\lambda \right) \rho}}-
\,{\frac {16\,{\lambda}^{4}-81\,{w}^{2}{\mu}^{2}}{9{\mu}^{2} \left(
3+2\lambda \right)^2 }}+\,{\frac {8{ \lambda}^{2} \left(
4\,{\lambda}^{3}+27\,{w}^{2}{\mu}^{2} \right) \rho }{9{\mu}^{3}
  \left( 3+2\lambda \right)^3 }} + {\cal{O}}(\rho^2) \right)
\chi(\rho) \\
\equiv \frac{d^2 \chi}{d \rho^2} + \left( \r ^{-1} \sum_{k\geq 1}
a_k \rho^{k-1} \right) \frac{d \chi}{d \rho} + \left( \r ^{-1}
\sum_{k\geq 1} b_k \rho^{k-1} \right) \chi(\rho)
 \end{multline}
 If $\chi$ is analytic around $\rho=0$:
\begin{equation}
\chi(\rho) = \sum_{k\geq 0} c_k \rho^{k},
\end{equation}
then (\ref{exp}) reduces to
\begin{equation} \label{sup}
s \left( s-1 + a_1\right) c_{{s}}+\sum _{l=0}^{s-1} \; c_{{l}} \;
\left( l \; a_{{s+1-l}} +  b_{{s-l}} \right)  = 0 , \;\; s=1,2,3,...
\end{equation}
In our case $a_1=-2$ so (\ref{sup}) splits into a system of three
equations in the unknowns $c_o,c_1$ and $c_2$ and an infinite system
on the $c_k, k>2$. Since the determinant of the finite system is
\begin{equation} \left|
 \begin{matrix} b_1 & a_1 & 0 \\ b_2 & b_1+a_2 & 2 a_1+2\\
 b_3 & a_3+b_2 & b_1+2 a_2 \end{matrix}  \right| = 0
 \end{equation}
 (for all $\mu,\lambda$ and $\omega$),
and the infinite system can be solved iteratively in terms of $c_3$,
(\ref{sup}) has a solution for arbitrary values of $c_o$ and $c_3$,
i.e., two linearly independent solutions (for more details on this
special case of singular point see e.g. \cite{braun}, pp.205-215)
$\Box$

\section{Static solutions}

Static solutions correspond to  $\omega=0$. In this case
(\ref{AppB2}) is reduced to,
\begin{equation}
\label{AppB3} {\frac {d^2 \chi}{d{r}^{2}}}
 -{\frac { \left(
4 \lambda {r}^{2}+6  \lambda r\mu -3{\mu}^{2} \right)  }{
 \left( 2  \lambda  r-3\mu \right)  \left( r+\mu
 \right) r}}{\frac {d \chi}{dr}} -{\frac { \left( 4{\lambda}^{2}{r}^{2}-6
\lambda  r\mu+3{\mu}^{2} \right)
  }{ \left( 2  \lambda  r-3\mu \right)
 \left( r+\mu \right) {r}^{2}}}{\chi} =0
\end{equation}
A straightforward computation shows that this equation has, in
general, two independent solutions, one regular and the other with
logarithmic divergences for $r=0$, and for $r=-\mu$.  The regular
solution is a polynomial of order $\ell+2$, given by,
\begin{equation}
\label{AppB4}
 Q(r)=\sum_{i=1}^{\ell +2} q_i r^i
\end{equation}
with
\begin{equation}
\label{AppB5}
  q_{{i}}={\frac { \left( i-2 \right) \Gamma  \left( {\ell}+i-1 \right)  \left[  \left( i-4 \right)
   {\ell}\, \left( {\ell}
+1 \right) +i-1 \right] } { 3\, \Gamma( {\ell}-i+3) {\mu}^{i -1}
\left[ \Gamma( i )  \right] ^{2}}}
\end{equation}
The general solution of (\ref{AppB3}) may be written in the form,
\begin{equation}
\label{AppB6} \chi(r) = C_1 Q(r) + C_2 Q(r) \int_{C_3}^r{\frac{y (2
\lambda y -3 \mu)}{(y+\mu)Q(y)^2} dy}
\end{equation}
where $C_1$, $C_2$ and $C_3$ are constants. Notice that this
expression contains effectively only two arbitrary constants,
because a change in $C_3$ amounts only to a change in $C_1$. We also
remark that, since all solution of (\ref{AppB3}) are regular for $r
\geq 0$, we must have $Q(r) > 0$ for $r>0$, for otherwise a zero
 in $Q(r)$ would introduce singularities in $\chi$ through the integral term in (\ref{AppB6}), which must be absent in the general solution of (\ref{AppB3}).

\subsection{Boundary conditions}

Equations (\ref{AppB5},\ref{AppB6}) provide the general time
independent solutions for the perturbative zero mode, but we still
need to specify boundary conditions. The question is then if these
solutions, perhaps for some particular choices of $C_1$ and $C_2$,
are acceptable as perturbations of the negative mass Schwarzschild
metric. As a possible criterion, we explore the change in a
particular curvature invariant, which we choose
 to be the Kretschmann invariant ${\cal{K}}$, given by,
\begin{equation}
\label{RW9} {\cal{K}} = R_{abcd} R^{abcd}
\end{equation}
where $R_{abcd}$ is the Riemann curvature tensor. Replacing
(\ref{AppB5})-(\ref{AppB6}) in the appropriate expression for
${\cal{K}}$, linearized to
 first order in the perturbations, we find that, in the $r \simeq 0$
 limit
\begin{multline}
\label{Kcerca} {\cal{K}} = \frac{12 \m^2}{r^6} - \epsilon \; C_2 \;
\left[  \frac{36 \m^2}{r^7}  + {\cal O} \left( \frac{\ln(r) }{r^5}
\right) \right] Y_{(\ell,m)}
\\ - \epsilon \; C_1 \;\left[ \frac{\ell (\ell ^2-1) (\ell +1)
(\ell+2)(\ell ^2+\ell +4)(\ell ^2+\ell-1)}{ 9 \; \m^2 r^3} + {\cal
O} \left( \frac{1}{r^2} \right)\right] Y_{(\ell,m)}
\end{multline}
 We must set $C_2=0$ to ensure that the first order term
does not grow faster than the zero order term as $r \to 0^{+}$. On
the other hand, if we  compute ${\cal{K}}$ for large $r$,  we obtain
\begin{multline}
\label{Klejos} {\cal{K}} = \frac {12 \mu^2} {{r}^{6}} -  \; \epsilon
\; C_1 \; \left[ \left( \frac{2 \; \Gamma (2 \ell +1)}{\mu^{\ell} \;
\Gamma (\ell-1)^2} \right) r^{\ell-5} +  {\cal O} (r^{\ell-4})
\right] \; Y_{(\ell,m)} \\
+ \epsilon \; C_2 \; \left[ \frac{K_{\ell}}{ r^{\ell+7}} +  {\cal O}
\left( \frac{1}{r^{\ell+8}} \right) \right] \; Y_{(\ell,m)},
\end{multline}
where
\begin{equation}
K_{\ell} = \frac{3 \sqrt{\pi} \; \Gamma(\ell)  (3 \ell^2+5 \ell
+8)(\ell+1)(\ell+2) \;\m^{\ell +2}}{ 4^{\ell} \; \Gamma(\ell +1/2)
(\ell-1)^2},
\end{equation}
then $C_1$ must also be set to zero in order to keep the
perturbation smaller than the zero order term  at large $r$.
 We must, therefore, conclude that there is
no (non trivial) choice of $C_1$ and $C_2$ for which a static
 perturbation will remain small in the entire range of $r$.\\

\subsection{A physical interpretation for the zero modes}

Even though there are problems in interpreting the zero modes given
by (\ref{AppB4})-(\ref{AppB6}) as
 perturbations of the negative mass Schwarzschild metric, they are static solutions
  of the linearized equations and, therefore, their physical interpretation is of interest.
  One way in which the problem may be analyzed is to consider the Schwarzschild metric in
   Weyl coordinates. The general form of the static axially symmetric vacuum metrics in Weyl
   coordinates is,
\begin{equation}
\label{wey1} ds^2=-e^{2U} dt^2+ e^{2k-2U} d\rho^2+\rho^2 e^{-2U}
d\phi^2 +e^{2k-2U} dz^2
\end{equation}
where $U=U(\rho,z)$, $k=k(\rho,z)$, and the Einstein equations
imply,
\begin{equation}
\label{wey2} \frac{\partial^2 U}{\partial \rho^2} +\frac{1}{r}
\frac{\partial U}{\partial \rho}+\frac{\partial^2 U}{\partial z^2}=0
\end{equation}
and,
\begin{eqnarray}
\label{wey3}
 k,_{\rho} & = & \rho \left(U,_{\rho}\right)^2-  \rho \left(U,_{z}\right)^2 \nonumber \\
k,_{z} & = & 2 \rho  U,_{\rho}U,_{z}
\end{eqnarray}
where a comma indicates partial derivative. It can be checked that
(\ref{wey2}) is the integrability condition for the system
(\ref{wey3}), so that $k$ is obtained by quadratures if $U$,
satisfying (\ref{wey2}) is given. The latter is the Laplace equation
in flat cylindrical coordinates and, as is well known, allows for
the construction of static vacuum solutions of the Einstein
equations by taking for $U$ different
 solutions of the Laplace equation corresponding to localized sources. The Schwarzschild space
 time is recovered by taking for $U$ the Newtonian potential corresponding to uniform line
  source of finite length, centered at the origin, and laying along the z-axis. The explicit
   forms for $U$ and $k$ in this case are,
\begin{eqnarray}
\label{wey4}
U_S(\rho,z) & = & \frac{1}{2} \ln\left[\frac{\rho_+ +\rho_- -2M}{\rho_+ +\rho_- +2M}\right] \nonumber \\
k_S(\rho,z)& = &  \frac{1}{2} \ln\left[\frac{(\rho_+ +\rho_-)^2-4M}{
4 \rho_+ \rho_-}\right]
\end{eqnarray}
where $\rho_{\pm}=\sqrt{\rho^2+(z\pm M)^2}$. The constant $M$ is
identified with the Schwarzschild mass and can be positive or
negative. Suppose now that instead of $U_S$ we consider, e.g.,
\begin{equation}
\label{wey5} U(\rho,z) = U_S(\rho,z)+U_1(\rho,z;m)
\end{equation}
where $U_1(\rho,z;m)$ is a solution of (\ref{wey2}) that depends
smoothly on a real parameter $m$ such that,
\begin{equation}
\label{wey6}
 U_1(\rho,z;0)= 0
\end{equation}
This implies that $U(\rho,z)$ given (\ref{wey5}) satisfies also
(\ref{wey2}) and can be used to construct static vacuum solutions of
the Einstein equations. But, in accordance with (\ref{wey6}), these
solutions, at least locally, can be made to approach the
Schwarzschild metric arbitrarily closely, by taking $m$ small
enough. This may fail, and in general it does,
 when we approach the line source of $U_S$. This line source corresponds to the regular horizon
 for $M>0$ and, therefore, a singularity in $U_1$ destroys that regularity, and $U_1$ cannot be
  considered globally as a ``small" perturbation. Similarly, for $M<0$, the line source
  corresponds to a naked singularity, and a singularity in $U_1$ may modify its structure,
  and can make it even ``harder''. To illustrate this discussion we consider for $U_1$ the form,
\begin{equation}
\label{wey7} U_1(\rho,z;m) = \frac{m}{\sqrt{\rho^2+z^2}}
\end{equation}
This represents (in Weyl coordinates) a point source at the origin.
The resulting equations
 for the modified $k(\rho,z)$ may be solved explicitly, for any $m$, but we are only interested
 in the change to order $m$. Actually what we are interested in is the form of the metric
 in Schwarzschild, rather than Weyl coordinates. The coordinate change that accomplishes
 this is well known (see, e.g. \cite{exactas}, for details). Performing it we find that
 in Schwarzschild coordinates the modified metric, to first order in $m$, takes the form,
\begin{eqnarray}
\label{wey8} g_{tt} & = & -1+ {\frac {2M}{r}}-{\frac {2 m \left(
r-2M \right) }{r \sqrt {{M}^{2} \cos^2 \theta   +{r}^{2} -2Mr}}}
\nonumber \\
g_{rr} & = & {\frac {r}{r-2 M}}+{\frac {2 mr \left( 2r-3M-2 \sqrt {
{M}^{2} \cos^2 \theta +{r}^{2} -2Mr} \right) }{ \left(r -2M\right)
\sqrt { {M}^{2}\cos^2 \theta    +{r}^{2}-2Mr}M}}
\nonumber \\
g_{\theta \theta} & = & {r}^{2}+{\frac {2 m {r}^{2} \left(2 r
-3M-2\sqrt {
 {M}^{2} \cos^2 \theta   +{r}^{2}-2Mr}
 \right) }{M\sqrt {M^2  \cos^2  \theta +{r}^{2}-2Mr}}}
\nonumber \\
g_{\phi \phi} & = &  r^2 \sin^2  \theta -{\frac {2 m r^2
  \sin^2  \theta  }{
\sqrt { M^2 \cos^2 \theta +{r}^{2} -2Mr}}}
\end{eqnarray}
This metric may be expanded now in the Regge-Wheeler-Zerilli
functions. For the metric in the form (\ref{wey8}), in the notation
of \cite{RW}, the only non vanishing functions are $H_0$, $H_2$,
$K$, and $G$, corresponding to even (scalar) perturbations for even
$\ell$, as the result of the  symmetry of $U_1(\rho,z;m)$ under $ z
\leftrightarrow -z$. This is not in the Regge-Wheeler gauge, but,
from gauge invariance \cite{moncrief}, the function $\psi(r,t)$ is
in this case given by,
\begin{eqnarray}
\label{wey9}
\psi(r,t) & = & \frac{1}{\lambda r+3M} \chi(r,t) \nonumber \\
\chi(r,t) & = & \frac{2r \left(r -2M\right)}{\ell (\ell+1) }  \left(
H_2   -r{ \frac {\partial K}{\partial r}}   -{\frac { \left( r- 3M
\right) K  }{r-2M}} \right)  + {r}^{2}  \left( K   +
 \left(r -2M \right) {\frac {\partial G}{\partial r}}    \right)
\end{eqnarray}
Going back to (\ref{wey8}), if we specialize to $\ell=2$, we find,
\begin{eqnarray}
\label{wey10} H_2 & = & {\frac {\sqrt { 5 \pi } m \left(
2r-3M\right)  }{2{M}^{4}}} \left[  \left( 2\,{M}^{2}-6
\,Mr+3\,{r}^{2} \right)  \ln  \left(1 -\frac{2M }{r}\right) +6\,
\left( r-M \right) M \right]
  \nonumber \\
K & = &  {\frac {\sqrt {5 \pi } m }{4{M}^{4}}}\left[ \left( 3\,{r}^{
3}-15\,{r}^{2}M+16\,r{M}^{2}-2\,{M}^{3} \right)   \ln  \left(1 -
\frac{2M}{r} \right) -24\,r{M}^{2}+10\,{M}^{ 3}+6\,{r}^{2}M \right]
\nonumber \\
G & = & {\frac {\sqrt {5 \pi }m}{12{M}^{4}}} \left[ 3 \left(M- r
 \right)  \left( 2{M}^{2}-2Mr+{r}^{2} \right)  \ln
 \left(1 -\frac{2M}{r} \right)
+2 M \left(6 r M -7 M^2 -3 r^2 \right)
  \right]
\end{eqnarray}
and the corresponding expression for $\chi(r)$ results,
\begin{equation}
\label{wey11} \chi= {\frac {  \sqrt {5 \pi } m r
}{6{\mu}^{3}}}\left[ 3 \left( 4{r}^{3}-6{r}^{2}\mu+3{\mu}^{3}
 \right)   \ln  \left( 1+\frac{\mu}{r}
 \right)  -12\mu {r}^{2}+24\mu^2 r-13{\mu}^{3}
 \right]
\end{equation}
which coincides with (\ref{AppB4})-(\ref{AppB6}) for  an appropriate
choice of integration constants. Similar results hold for the other
even values of $\ell$. Non vanishing contributions for odd $\ell$
may be
 obtained starting from a different choice of $U_1(\rho,z;m)$ that
 makes it odd under $z \longleftrightarrow -z$, with analogous results
 regarding the corresponding zero modes. Therefore, the zero modes are the linearized
expressions for the modifications that result in the Schwarzschild
metric by the addition of different (static) multipoles. We stress
again that these modifications approach only locally the
Schwarzschild metric, because they modify non trivially the
structure of its singularity. We shall add more comments on this
feature in Section \ref{cc}.

\section{Time dependent solutions}

Going back to (\ref{AppB2}), we consider now the general, time dependent case, with
 $\omega \neq 0$. We are interested in solutions that are regular in $0\leq r \leq \infty$.
 Therefore, we need to consider the behavior of $\chi_1$ near the singular points
  $r=0$, $r=3\mu/(2 \lambda)$ and $r=\infty$. First we find that near $r=0$,
  any solution of (\ref{AppB2}), for any $\omega$, is of the form,
\begin{equation}
\label{RW20} \chi(r) \simeq C_1 r + C_2 r \ln(r) + \dots
\end{equation}
where $C_1$ and $C_2$ are constants, and $\dots$ indicates terms
that vanish faster for $r=0$. This implies that all solutions of
(\ref{AppB2}) vanish for $r=0$, but, in general, $d \chi/dr$
diverges as $\ln(r)$. On account of our previous discussion
regarding acceptability of the perturbations, this means that we
need to impose the condition that this divergence is absent, and
therefore, only the regular solution associated with $C_1$ can be
considered for the stability analysis. This fixes our boundary
condition for $r=0$. The behavior at $r=\infty$ is also simple. For
large $r$, the general solution of (\ref{AppB2})
 is of the form (\ref{coso7}) \footnote{ Notice that the leading term in $r$ in
(\ref{coso7}) can be traced to the
 transformation (\ref{AppB1}), since, for any sign of $M$, the Zerilli equation for $\psi(t,r)$
 approaches the one dimensional wave equation, and $\psi(t,r) \rightarrow \exp(i \omega
 t)$.}. We remind the reader that, in view of Lemma 2 in Section
 \ref{rwzg}, $\chi(r)$ is analytic at $r= 3 \mu / (2 \lambda)$

From the assumed time dependence, positive values for $\omega^2$
correspond to an oscillating, and therefore stable, behavior for the
perturbations. The interesting range for instability is, therefore,
$\omega^2 < 0$.
 On account of (\ref{RW17}), the solutions for $\omega^2 < 0$ will
grow unboundedly for large $r$, unless we can eliminate the exponentially divergent terms.
  In the following section we analyze this problem in detail.

\section{Unstable solutions} \label{us1}

Unstable modes, correspond to solutions of (\ref{AppB2}) with $\omega^2 =-k^2$, that
 satisfy the regularity condition for $r=0$, and are exponentially decreasing for large $r$.
  Therefore, finding these solutions  turns this equation into a boundary value problem,
  determining the appropriate value or values of $k$, if they exist.

As a first step, we considered a ``shooting'' approach, and, using a high precision
numerical method, we found, for $\ell=2$, a numerical solution for $k \simeq 4/\mu$.
Since this value was so close to an integer times $1/\mu$, we
replaced $k = 4/\mu$, $\ell=2$ in (\ref{AppB2}), and found that indeed it has a rather
simple solution for this value of $k$. Repeating the procedure for $\ell=3$, we found again
a solution for $k=20/\mu$. This suggested the possibility that we could construct solutions
 for {\em all} $\ell$, by a simple generalization of the results for $\ell=2$, and $\ell=3$.
 This was effectively the case, and we found that,
\begin{equation}
\label{AppC3}
 {\chi} ( r )  = r{e^{-kr}} \left( r+\mu \right) ^{k\mu}
 \left[ {C_1}+{C_2}\,\int _{{C_3}}^{r}\!{\frac { \left( 2\,
\lambda\,y-3\,\mu \right) ^{2}{e^{2\,ky}}}{ \left( y+\mu \right) ^{2\,
k\mu+1}y}}{dy} \right]
\end{equation}
where $C_1$, $C_2$, and $C_3$ are arbitrary constants, is a solution of (\ref{AppB2}) with
$\omega^2=-k^2$, provided  the constant $k$ is given by
\begin{equation}
\label{AppC4}
 k = \frac{2\lambda (\lambda+1)}{3 \mu}=\frac{ (l-1) l (l+1) (l+2)} {6 \mu}
\end{equation}
and, therefore, $k\mu$ is an integer for all $\ell$. Notice that (\ref{AppC3}) effectively
 contains only two arbitrary constants, because any change in $C_3$ can be compensated by a
 corresponding change in $C_1$.

It is easy to see that the integral in (\ref{AppC3}) contributes logarithmic divergences for
 $r=0$, and an exponentially increasing factor for $r \rightarrow \infty$. Therefore if we
 choose $C_2=0$, we find that
\begin{equation}
\label{AppC5}
 {\zeta} ( t, r )  = C_1 r{e^{-kr}} \left( r+\mu \right) ^{k\mu}e^{kt}
\end{equation}
with $k$ given by (\ref{AppC4}), is a solution of (\ref{AppB2}), representing
 the general expression for the regular, bounded, unstable modes.
Moreover, this result implies that {\em all scalar modes (i.e. for all $\lambda$) of the Schwarzschild negative
mass naked singularity are linearly unstable}.

As a check of the consistency of our results, we replace (\ref{AppC5}) in (\ref{AppB1}),
 and this in(\ref{RW2}), and obtain
\begin{eqnarray} \label{us}
\label{RW24a}
 K(t,r) & = & -\frac{ (\lambda + 1)(r+\mu)^{k \mu}}{3 \mu}\exp[k(t- r)]   \\
 H_1(r,t) & = & -H_2(t,r) = -\frac{\lambda (\lambda+1)
   [2(\lambda+1) r+ 3 \mu] \; r  (r+\mu)^{k\mu -1}}{9 \mu^2}\exp[k(t- r)] \nonumber
\end{eqnarray}
which shows explicitly the regularity of the solution of the perturbative equations in the
 range $0 \leq r < +\infty$. The perturbations of the curvature are also well behaved.
Replacing in the Kretschmann scalar, we find that it is unchanged from the unperturbed value,
to order $\epsilon$.
Furthermore, a complete set of quadratic algebraic invariants of the
Riemann tensor \cite{cm,zm} was computed to find  that they are all
invariant to first order in $\epsilon$. However, we have found non
trivial first order contributions to the  {\em differential}
invariants
\begin{eqnarray} \label{i1}
R_{abcd;e} \; R^{abcd;e} &=& \frac{180 \mu^2 (r + \mu)}{r^9} - \e \;
  \frac{ 5 \ell (\ell-1) (\ell+1)(\ell+2)}{r^8}  \;  Y_{\ell m}(\theta,\phi)\\
\nonumber && \times  \left[ r \ell (\ell+1) + 3 \mu \right]
  (r+\mu)^{k \mu} \;
 \exp[k (t-r)]
\end{eqnarray}
and \begin{multline}
 \label{i2}
R_{abcd;ef} \; R^{abcd;ef} = \frac{90 \m^2 (65 \m^2 +120 \m r+56
r^2)}{r^{12}} \\- 30 \e k \frac{(r+\mu)^{k \mu} \exp[k
(t-r)]}{r^{11}} \; Q_{\ell}(r) \;Y_{\ell m}(\theta,\phi)
\end{multline}
\begin{multline}
Q_{\ell}(r) = \left( {l}^{2} \left( l-1 \right)  \left( l+2 \right)
\left( l+1 \right) ^{2 }{r}^{3}+3\,\mu\,l \left( l+1 \right)  \left(
{l}^{2}+l+14 \right) {r} ^{2} \right. \\ \left.+ 3 \,{\mu}^{2}
\left( 17\,{l}^{2}+17\,l+56 \right) r+180\,{\mu}^{3} \right)
\end{multline}
and we may expect non trivial contributions in some of the many other differential
invariants. Note that, both for  (\ref{i1}) and (\ref{i2}),
the ${\cal O} (\e)$ term  grows slower than the ${\cal O} (\e^0)$ term as
$r \to 0^+$ limit, and also as $r \to \infty$.

\section{Comparison with  previous results} \label{sean}

In \cite{ghi}, the Hilbert space
\begin{equation} \label{hs}
{\cal H} = L^2(r^*,dr^*)
\end{equation}
is introduced together with Zerilli's equation (\ref{RW3b}). Since
the LHS of (\ref{RW3b}) is a self adjoint operator in ${\cal H}$, it
admits a basis of eigenfunctions that allow to evolve any initial
data $(\psi, \partial_t \psi)|_{t=0} \in  {\cal H} \otimes {\cal
H}$. It was also shown in \cite{ghi} that (for $\ell = 2$) there
is a $\w^2= 0$ solution  $\hat \psi_o \in {\cal H}$ of (\ref{RW3b}).
It is important to review at this point how $\hat \psi_o$ is
constructed. For negative $M = -\m /2$ the $\ell=2$ Schr\"odinger
potential
\begin{equation}
V = 2 \left( 1+\frac{\m}{r} \right) \frac{4 r^2 \left(3 r -3 \m/2
\right) + \frac{9}{4}\m^2 \left(2 r - \m/2 \right) }{ r^3 \left( 2 r
- \frac{3 \m}{2} \right) ^2 },
\end{equation}
is singular at $r_c \equiv 3 \m /4$. The general solution of
(\ref{RW3b}) for $\omega=0$ \cite{ghi}
\begin{multline}
\psi_o = C_3 \frac{r (3 \m^3- 6 r^2 \m + 4 r^3)}{\m ^3 (4r -3 \m)}
\\+ C_4 \left[\frac{r (13 \m^3 +  12 r^2 \m - 24 \m^2 r)}{3 \m ^3 (4r
-3 \m)} - \frac{r (3 \m^3- 6 r^2 \m + 4 r^3)}{\m ^3 (4r -3 \m)} \log
\left( \frac{r+\m}{r} \right) \right]
\end{multline}
lies outside ${\cal H}$, however, for $C_3 /C_4 = \log(7/3)  - 4/9
\equiv q_o$, both $\psi_o$ and its first derivative vanish at $r_c$
\cite{ghi}, then $\psi_o$ and can be matched at $r_c$ to the trivial
$\w=0$ solution giving
\begin{equation} \label{nosol}
\hat \psi_o (r) = \begin{cases} \psi_o(r) & , 0 \leq r < r_c  \\
0 & , r > r_c \end{cases}
\end{equation}
Note that $\hat \psi_o$ is an $\w=0$ solution in ${\cal H}$, whose
second derivative has a discontinuity at $r=r_c$. The existence of
such a solution signals a critical value of $q \equiv C_3/C_4$
separating unstable ($\omega^2 < 0$)
from stable solutions {\em for the Zerilli equation in ${\cal H}$}.\\
In the previous section we exhibited a $C^{\infty}$ perturbed metric
which is an unstable solution of the linearized Einstein's vacuum
equation around a negative mass \s spacetime, and showed that it
satisfies appropriate boundary conditions, both at the singularity
and at infinity. However, we have not  given a procedure to evolve
arbitrary perturbations. The reason why we have not adopted the
Hilbert space (\ref{hs}) is that, although (\ref{hs}) is a
natural choice in other contexts, generic linear perturbations
(\ref{RW1}) of the metric of a negative mass \s spacetime lye
outside ${\cal H}$, since the integral
\begin{multline}
(\psi_Z,\psi_Z) = \int _{0}^{\infty} |\psi_Z|^2 \frac{dr}{(1+\m/r)}
\\ = \int _{0}^{\infty} \left| \frac {r (r+\m)}{ 6  ( r - 3\m /4 ) }
\left(H_2
 - r  \frac {\partial K}{\partial r}
\right)
    +\frac{r}{3} K \right|^2 \; \frac{dr}{(1+\m/r)}
\end{multline}
diverges due to the pole at $r= r_c = 3 \mu /4$ (here we have
specialized to $\ell=2$ and used (\ref{RW3a}) to write the
$L^2(r^*,dr^*)$ norm of the perturbation in terms of the perturbed
metric elements, which may be arbitrary  $C^2$ functions for $r \in
(0,\infty)$).  Even compactly supported metric perturbations are
excluded as initial data in the formalism developed in \cite{ghi},
unless they meet the rather artificial requirement that, for every
$\ell \geq 2$, its $\ell-th$ harmonic projection satisfies
$$  \left(H_2^{\ell}
 - r  \frac {\partial K^{\ell}}{\partial r}
\right) \left|_{\left(r = \frac{3 \mu }{2 (\ell-1) (\ell+2)}\right)}
\right. = 0
$$ The Zerilli function $\psi_Z$ is extremely useful because it
reduces a complicated PDE system to a much simpler differential
equation. However, care must be exercised when deciding what kind of
solutions of this differential equation one is interested in, since
$\psi_Z$ has, for the negative mass \s spacetime, a built in
singularity, as seen, e.g., from (\ref{RW3a}). As an example, it
follows from (\ref{RW1}) and (\ref{RW2}), that the perturbed metric
from (\ref{nosol}) has a discontinuity  at $r=r_c$. In particular,
it will not give a linearized solution of the vacuum Einstein's
equations, in spite of being a solution of Zerilli's equation. This
implies that, although (\ref{nosol}) gives a marginally stable
solution for the Zerilli problem in ${\cal H} = L^2(r^*,dr^*)$, it
does not prove the existence of a  marginally stable solution of the
linearized Einstein's equations we are dealing with, since only
$C^4$ solutions to the Zerilli equation are relevant to the
linearized gravity problem. Our choice of initial data set is that
$\zeta|_{t=0}$ and $\p \zeta / \p t |_{t=0}$ be $C^{\infty}$
functions subject to (\ref{cz}). As follows from Lemma 1 in Section
\ref{rwzg}, this is equivalent to requiring that the metric
perturbation be $C^{\infty}$. Equation  (\ref{RW17}) is a hyperbolic
differential equation in $(t,r)$ from which the evolution of given
initial data for $H_1$, $H_2$ and $K$ can be computed. The
singularities in the coefficients of $\partial \chi/\partial r$ and
$\chi$, for $2 \lambda r = 3 \mu$ were shown to be irrelevant in
Lemma 2 of Section \ref{rwzg}.
  The
hyperbolicity of (\ref{RW17}), (or the regularity and smoothness of
the light cones for the background metric), imply that smooth
initial data of compact support given at, say, time $t=t_0$, should
evolve into initial data also of compact support for $t=t_1$, at
least for sufficiently small $t_1-t_0$, then compactly supported
smooth $\zeta|_{t=t_o}$ and $\p \zeta / \p t |_{t=t_o}$ satisfying (\ref{cz})
evolves into a compactly supported $\zeta|_{t=t_1}$\\
The foregoing discussion shows that  (\ref{us}) can actually be
excited without a fine tuning of the initial condition. We look at
solutions of (\ref{RW17}) corresponding to a time dependence of the
form $\exp(i \omega t)$, with $\omega$ {\em arbitrary complex},
without the logarithm singularity ($C_2=0$ in (\ref{RW20})). A
higher order expansion around $r=0$
\begin{equation} \label{coso6} \tilde{\chi}(r,\omega) \propto \left[r
- \frac{\lambda(1+\lambda)}{3 \mu^2} r^3+
\frac{2\lambda(1+\lambda)}{9 \mu^3} r^4 + {\cal{O}}(r^5)\right]
\end{equation}
reveals  that there is no term in $r^2$, and only the coefficients
of $r^5$ and higher order depend on $\lambda$ and on (even) powers
of $\omega$. We have already shown in Section \ref{rwzg} (Lemma 2)
that all solutions of (\ref{RW17}) for this type of time dependence
are smooth at $r = 3 \mu /(2 \lambda)$. We may similarly obtain an
asymptotic expansion for these solutions for $r\rightarrow \infty$.
These can be written in the form,
\begin{equation} \label{coso7} \tilde{\chi}(r,\omega)  \propto  e^{[i k
(r-\mu \ln(r+\mu)]} \left[r - \frac{2 i \lambda(1+\lambda) -3 k
\mu}{2 k \lambda}   +\frac{2 \lambda(1+\lambda) + 3 i k \mu}{4
\omega^2 r} +{\cal{O}}(1/r^2)\right] \end{equation} where $k = \pm
\omega$, and the remaining coefficients depend only on even powers
of $\omega$. For real $\omega$, and large $r$, these solutions
represent (modulated) incoming and outgoing waves, depending on the
sign of $k/\omega$. Since (\ref{RW17}) is linear, any linear
superposition of solutions is also a solution. In particular, we may
consider superpositions of the form,
\begin{equation} \label{coso8}
\chi(r,t)=\int_{-\infty}^{+\infty}{{\cal{K}}(\omega) e^{i \omega
t}\tilde{\chi}(r,\omega) d\omega} \end{equation} which provide
everywhere regular solutions (including $2\lambda r=3\mu$, for a
wide variety of functions ${\cal{K}}(\omega)$. An interesting choice
is, \begin{equation} \label{coso9} {\cal{K}}(\omega)= C
e^{-(\omega-\omega_0)^2  q^2/4} \end{equation} where $C$,
$\omega_0$, and $q$ are constants. Then, for large $q$, we may use a
steepest descent estimation of (\ref{coso8}) to find that for
${\cal{K}}(\omega)$ of the form (\ref{coso9}), near $r=0$ we have,
\begin{equation} \label{RW170} \chi(r,t) \simeq K_1 e^{-t^2/q^2}
e^{i \omega_0 t}\left[r - \frac{\lambda(1+\lambda)}{3 \mu^2} r^3+
\frac{2\lambda(1+\lambda)}{9 \mu^3} r^4 + {\cal{O}}(r^5)\right]
\end{equation} where $K_1$ is a constant, and terms of order $r^5$
or higher are polynomial in $t$. This result implies that the
amplitude near $r=0$ is strongly suppressed for large $t$. For large
$r$, on the other hand, we find, \begin{eqnarray} \label{RW171}
\chi(r,t) & \simeq & K_2 \exp[i \omega_0
u_{\pm}]\exp[-(u_{\pm})^2/q^2] \nonumber \\ & & \times \left[r -
\frac{2 i \lambda(1+\lambda) -3 k \mu}{2 k \lambda}   +\frac{2
\lambda(1+\lambda) + 3 i k \mu}{4 \omega^2 r}
+{\cal{O}}(1/r^2)\right] \end{eqnarray} where $K_2$ is constant,
and, \begin{equation} \label{RW172} u_{\pm}(r,t)= t \pm
[r-\mu\ln(r+\mu)] \end{equation} Since this expansion is valid only
for large $r$, we find that solutions corresponding to $u_+$ are
strongly suppressed for all $t$, while those for $u_-$ represent an
outgoing wave packet, roughly centered at $t = [r-\mu\ln(r+\mu)$,
moving along a null geodesic of the unperturbed spacetime. All these
solutions represent the evolution of some smooth initial data that
can be obtained from (\ref{coso8}), for $t=0$. Clearly, these
superpositions display a "stable" behaviour, since they do not
contain modes that grow exponentially in time. We notice, however,
that we may combine any of these solutions with an arbitrarily
small, but nonzero contribution from the unstable mode given by eqs.
(\ref{AppC5})-(\ref{RW24a}), to obtain an equally smooth initial
data at $t=0$, whose evolution displays the exponentially growing
mode. We thus find a very large set of smooth initial data leading
to an equally smooth ``unstable" behaviour, showing that such data
need not be
particularly ``fine tuned".\\

Another issue discussed in \cite{ghi} is that of the gravitational
energy of the perturbation. Here again we find a number of
subtleties, apparently peculiar to the negative mass case. The
unstable solution (\ref{us}) lies outside ${\cal H}$ in \cite{ghi},
and therefore outside the scope of ref \cite{ghi}. To no surprise,
its ``Zerilli energy"
\begin{equation} \label{ez}
E_Z := \int \left( \left| \frac{\p \psi_Z}{\p t} \right| ^2 + \left|
\frac{\p \psi_Z}{\p r^*} \right| ^2 + V | \psi_Z |^2 \right) dr^*
\end{equation}
diverges. It is argued in \cite{ghi} that $E_Z$ agrees with the
notion of gravitational energy of a perturbation from second order
perturbation theory, i.e.,
\begin{equation} \label{sean1}
E_G = - \frac{1}{8 \pi} \int G^{(2)}_{ab} \eta^a \zeta^b d
\Sigma_{(3)},\hspace{1cm}\eta = f^{-1/2} \partial / \partial t,
\;\;\; \zeta = \partial / \partial t,
\end{equation}
$G^{(2)}_{ab} $ being the second order correction to the Einstein
tensor. However, from (\ref{us}) we get, for the $(\ell,m=0)$ mode,
\begin{equation} \label{sean2}
 G^{(2) \ell}_{tt} = \exp(kt) \left[ A(r,\ell) \left( Y_{(\ell,0)} \right)^2 +
 B(r,\ell) \left( \frac{\partial Y_{(\ell,0)}}{ \partial \theta}
 \right)^2 \right]
 \end{equation}
where  $A$ and $B$ are  $C^{\infty}$ functions of $r$ in $(0,
\infty)$, falling off exponentially as $r \to \infty$, and behaving
as $c/r^3$ for $r \simeq 0$. This implies that both
$$\int_0^{\infty} \frac{r^2 A}{f} dr \hspace{1cm}\text{and} \hspace{1cm}
\int_0^{\infty} \frac{r^2 B}{f} dr$$ converge,
 and thus $E_G$ is finite for the  perturbation (\ref{us}),  showing that $E_Z \neq E_G$
 in the negative mass case. \\
 We may try to find an alternative Hilbert space and energy concept
 associated to a
Zerilli-like equation
\begin{equation}
\frac{\p^2 \phi}{\p t^2} + \o (\phi)= 0, \hspace{1cm} \o = A(r)
\frac{\p^2}{\p r^2} + B(r) \frac{\p}{\p r} + C(r).
\end{equation}
Yet, is not hard to see that, for unstable spacetimes, all unstable
modes will either be outside the Hilbert space, or have zero energy.
To see this suppose we choose {\em any} hermitian product $<,>$ in
an appropriate function space under which $\o$ is self-adjoint:
$<\phi_1,\o \phi_2> = <\o \phi_1,\phi_2>$. Note that the  hermitian
product may  well involve $\phi$ and $r-$derivatives of $\phi$.
Adding  $0 = <\phi_t,\phi_{tt} + \o \phi>$ to its complex
conjugated, and using the self adjointness of $\o$, we arrive to a
conserved ``energy"
\begin{equation} \label{en}
\frac{d E}{d t} \equiv \frac{d}{dt} \left[ <\phi_t,\phi_t> + <\phi,
\o\phi> \right] = 0
\end{equation}
This  is essentially a generalization of the argument used in
\cite{chandra} to obtain the conserved energy (\ref{ez}) above. We
would like to use it now to prove that, if the underlying spacetime
is unstable,  the unstable modes will either be non normalizable
(and then $E$ in (\ref{en}) undefined, as happens with the energy
$E_Z$ for (\ref{us})), or will trivially vanish. In fact, suppose
there is a solution $\phi(t,r) = \exp(\kappa t) \phi_o(r), \kappa
>0$ and assume $\phi_o$ normalizable under $<,>$, so that the energy in
(\ref{en}) is well defined. Since
$$ E(t) = e^{2 \kappa t} \left[ \kappa^2 < \phi_0,\phi_0> + <\phi_0, \o \phi_0>
\right]$$ does not depend on $t$, it has to vanish.

\section{Comments and conclusions} \label{cc}

In this article we have shown that even if we only allow for the
 most restrictive boundary conditions, both as $r \to 0^+ $ and $r \to \infty$, the
 perturbation equations for the negative mass Schwarzschild metric contain,
 for every angular mode, solutions that grow exponentially in time.
 These solutions represent everywhere regular perturbations that can be
 made initially globally arbitrarily small and, therefore, imply that
 the metric is unstable at the perturbative level, for all angular modes.
 An interesting technical point of our derivations was the introduction of
a modification of the standard Regge-Wheeler-Zerilli method, that led to equation
 (\ref{AppB2}) and its rather unusual properties.
Our analysis of the zero modes indicates that rather than
perturbations they should be considered as linearized approximations
for exact static metrics that locally, but not globally approach the
negative mass Schwarzschild metric. Although asymptotically flat,
they actually contain a naked singularity where the curvature
diverges even faster than for the Schwarzschild metric. From the
perspective of the present analysis, these spacetimes should also be considered
unstable, since, if they are considered as part of the unperturbed
metric, one would be allowed to relax the boundary conditions at
$r=0$, and as can be seen from the derivations in Section \ref{us1}, this
would lead to unstable solutions for {\em any} $k = \sqrt{-\omega^2} $. Our
results also suggest that, even though the vector (linearized)
perturbations have been shown to be stable \cite{ghi}, one would
expect that these would also become unstable through the coupling to
the scalar modes. Since vector modes are related to rotation, this
again suggests that rotation would not make the singularity more
stable, although going through this in detail would require higher
order perturbation theory, or, perhaps, some application of the
Teukolsky equation.

Finally, there is an obvious implication of our results to the
cosmic censorship conjecture, since the linear instability would imply
that the negative mass \s cannot be the final stage of
an evolving spacetime.

\section*{Acknowledgments}
We acknowledge fruitful correspondence with and comments  from G.
Gibbons, A. Ishibashi and S. Hartnoll that helped us clarify the
main differences between their approach in \cite{ghi} and ours.
 G.D. thanks the KITP at UCSB, where part of this work was done,
 for  hospitality.
This work was supported in part by grants of CONICET (Argentina) and
the Universidad Nacional de   C\'ordoba. It was also supported in
part by grant NSF-INT-0204937 of the National Science Foundation of
the US. The authors are supported by CONICET (Argentina).

\end{document}